# The Greenland Telescope (GLT): Antenna status and future plans.

Philippe Raffin[*a], Juan Carlos Algaba-Marcos[a], Keichi Asada[a], Raymond Blundell[b], Roberto Burgos[b], Chih-Cheng Chang[a], Ming-Tang Chen[a], Robert Christensen[d], Paul K. Grimes[b], C.C. Han[a], Paul T.P. Ho[a,b], Yau-De Huang[a], Makoto Inoue[a], Patrick M. Koch[a], Derek Kubo[c], Steve Leiker[b], Ching-Tang Liu[e], Pierre Martin-Cocher[a], Satoki Matsushita[a], Masanori Nakamura[a], Hiroaki Nishioka[a], George Nystrom[a], Scott N. Paine[b], Nimesh A. Patel[b], Nicolas Pradel[a], Hung-Yi Pu[a], H.-Y. Shen[a], William Snow[c], T.K. Sridharan[b], Ranjani Srinivasan[c], Edward Tong[b], and Jackie Wang[a].

[a]Academia Sinica Institute of Astronomy & Astrophysics, P.O. Box 23-141, Taipei 10617, Taiwan
[b]Harvard-Smithsonian Center for Astrophysics, 60 Garden St, MS 78, Cambridge, MA 02138, USA
[c]Academia Sinica Institute of Astronomy & Astrophysics, 645 N A'Ohoku Pl, Hilo, HI 96720, USA
[d]Smithsonian Astrophysical Observatory, 645 N A'Ohoku Pl, Hilo, HI 96720, USA
[e]ASRD/CSIST N.300-5, Lane 277, Xi An Street, Xitun District,Taichung City 407, Taiwan

## ABSTRACT

The ALMA North America Prototype Antenna was awarded to the Smithsonian Astrophysical Observatory (SAO) in 2011. SAO and the Academia Sinica Institute of Astronomy & Astrophysics (ASIAA), SAO's main partner for this project, are working jointly to relocate the antenna to Greenland to carry out millimeter and submillimeter VLBI observations. This paper presents the work carried out on upgrading the antenna to enable operation in the Arctic climate by the GLT Team to make this challenging project possible, with an emphasis on the unexpected telescope components that had to be either redesigned or changed. Five-years of inactivity, with the antenna laying idle in the desert of New Mexico, coupled with the extreme weather conditions of the selected site in Greenland have it necessary to significantly refurbish the antenna. We found that many components did need to be replaced, such as the antenna support cone, the azimuth bearing, the carbon fiber quadrupod, the hexapod, the HVAC, the tiltmeters, the antenna electronic enclosures housing servo and other drive components, and the cables. We selected Vertex, the original antenna manufacturer, for the main design work, which is in progress.

The next coming months will see the major antenna components and subsystems shipped to a site of the US East Coast for test-fitting the major antenna components, which have been retrofitted. The following step will be to ship the components to Greenland to carry out VLBI and single dish observations. Antenna reassembly at Summit Station should take place during the summer of 2018.

## 1. INTRODUCTION

The Greenland Telescope (GLT) will be used for submillimeter VLBI observations and single dish terahertz (THz) observations in an extremely cold atmospheric environment. Submillimeter VLBI observations are detailed by Inoue et al. [1] and Single dish observations and instrumentation by Grimes et al [2]. Table **1** shows the typical characteristics of the site for the GLT operating conditions. Additional information is available at the website of Summit Station. [3]

|  | Primary operating conditions | Secondary operating conditions | Survival |
|---|---|---|---|
| Ambient temperature | 0 to -50 °C | 0 to -55°C | -73 °C |
| Vertical temperature gradient | 1 K.m$^{-1}$ | | n/a |
| Wind | 0 to 11 m.s$^{-1}$ | 0 to 13 m.s$^{-1}$ | 55 m.s$^{-1}$ |

Table 1: Environmental conditions at the GLT site

* Send correspondence to P. Raffin, e-mail: raffin@asiaa.sinica.edu.tw

# 2. ANTENNA RECOMMISSIONNING

## 2.1 Goal

The ALMA-NA Prototype antenna had been out of operation for about five years, at the Very Large Array (VLA) site in New Mexico. We contracted VERTEX Antennentechnik GmbH, Germany, the original designer of this antenna, to perform inspections and tests during the summer of 2011 to determine the status of the antenna after SAO received the antenna from NSF. In addition to determine the status of the antenna, these inspections and re-commissioning tests gave us a first idea of which components could be used in an Arctic environment and which components would have to be replaced or refurbished.

## 2.2 Antenna Inspection

All main reflector panels were inspected, and apart from a damaged panel, and uneven gaps between panels, they are fine. After the antenna was disassembled the following year, we discovered that one quadrupod leg was cracked and that most backup structure segments presented defects and damages. The receiver cabin equipment was functional (shutter, ventilation) however the wall temperature regulation system was not in good condition, the piping showed leakage and corrosion, indicating that a general refurbishment would be necessary. The elevation components, the yoke and the support cone appeared in good condition, with only some grease contamination. Drives and bearing raceways were lubricated. No anomalies were found and the drives operated properly after re-commissioning of the servo system. All fixation bolts were tested for proper tightening torque. Some cables in the azimuth cable wrap were not in proper position and were realigned before re-activation of the azimuth drive. All the other drive components were working properly. The azimuth encoder drive tube alignment was not checked, and the azimuth axis tiltmeter absent. Some motor cable connections were not in perfect condition and bolts showed corrosion.

The servo drive cabinet needed a complete cleaning, some relays behaved unreliably, and the air conditioning was not functioning properly, probably due to contamination by dust. PLC apart from the battery was still working correctly. The servo amplifiers were functional as well as all hard and soft limits, stow pin mechanism, interlock system, emergency stops and membrane shutter. During re-test one azimuth motor was replaced due to the non-working brake.

A complete cleaning of the 19" computer outdoor rack was carried out. The antenna control unit (ACU) was fully functional. However the pointing computer (PTC) was slow to boot and the CAN communication to the subreflector bus station was not stable. The optical encoders were found functional. Metrology sensors appeared to work properly, when present. The status of the hexapod could not be verified visually. After booting and initialization of the PTC, it was possible to move the subreflector with the hexapod, but further tests and upgrade would be necessary. The cabin HVAC system was no longer functional.

The electrical system, both the main power distribution in the support cone and the UPS needed a complete cleaning, cabinets were dirty inside and outside, batteries completely discharged, the voltage gauges showed signs of overheating. It was recommended to upgrade distribution boxes according to the latest regulations of the International Electro technical Commission (IEC).

## 2.3 Antenna re-Tests

The scope of re-tests consisted in re-calibration of the azimuth encoder, measurement of antenna friction, measurement of gearbox parameters, determination of antenna unbalance, measurement of tracking performance, and measurement of metrology sensor behavior over azimuth.

Azimuth friction as a function of position corresponded to the 2003 measurements and was quite smooth over the entire azimuth range. The friction as a function of velocity was better than previous measurements. In elevation, the friction as a function of position showed similar behavior as in 2003 except at positions above 90-degree in elevation, while the friction as a function of velocity was also lower than measured in 2003. VERTEX found only a slight change in the elevation unbalance compared to 2003 and assumed it came from the empty cabin or limited accuracy of the measurements.

Gearbox backlash and rigidity were as expected. Friction of the 2 azimuth gearboxes was slightly higher than the ALMA

production antenna specifications. In elevation, friction of the 2 gearboxes on one side of the antenna was within specifications whereas it was twice the specified value on the other side. This is probably an alignment problem and VERTEX recommended reworking the corresponding gear rim before re-assembly in Greenland.

Finally the tiltmeters, the linear sensors and their carbon fiber supporting rods were functioning properly and measurements did not show unexpected behaviors.

**2.4 Photogrammetry.**

Two photogrammetry measurements were done on the antenna to verify the reflector surface accuracy. New targets were installed on the aluminum panels and the V-Stars system from Geodetic Systems, Inc. (GSI) was used. The panels of the primary mirror had been untouched since 2006.

A first measurement (Figure **2**) was carried out at an elevation of 5° and the second measurement at 90° of elevation. Both measurements showed almost the same situation with a clearly visible astigmatism. The surface accuracy for measurement #1 was 78 μm (EL = 5°) and for measurement #2 (EL =90°) 79 μm. The elevation position of the dish has no influence on the surface accuracy of the antenna, given the high stiffness of the reflector.

Figure **1** shows the result of the measurement by holography done by Symmes et al [4] in 2006. The astigmatism component amounts to 249 μm peak-to-peak. The map of Figure **2** shows an astigmatism component of about 300 μm peak-to-peak, which is quite close to the astigmatism encountered 5 years before.

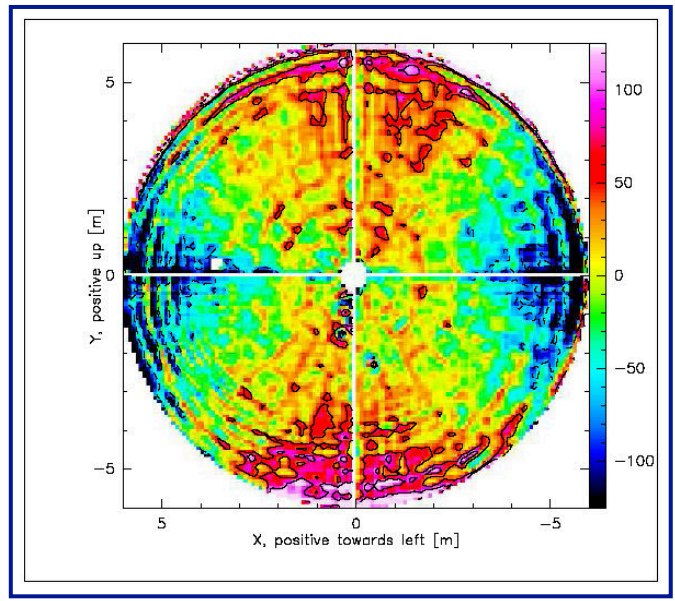

Figure 1: Average of 3 Maps at 104 GHz; holography scan in Elevation

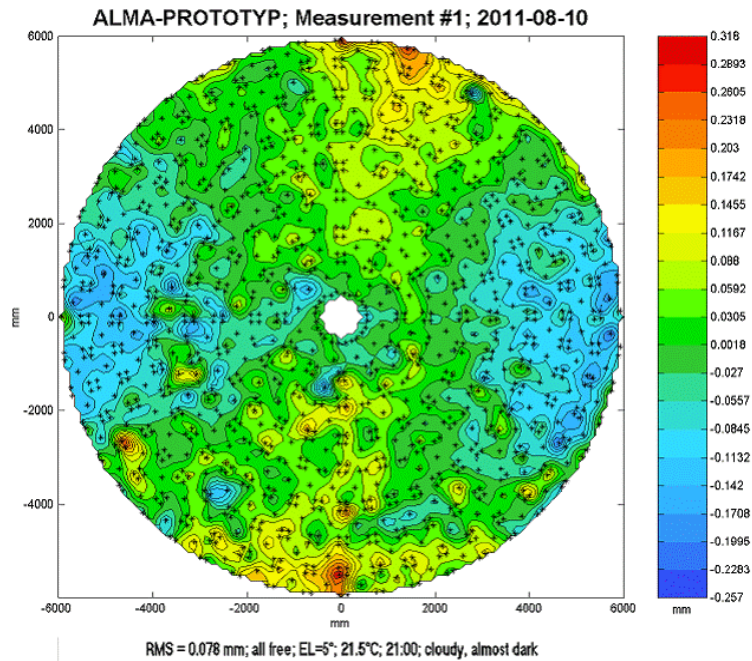

Figure 2: First photogrammetry measurement on the antenna, panel unmoved since 2006. Astigmatism is clearly visible.

After a complete panel adjustment session, lasting almost 3 days, the surface was again measured at three different elevations. All three measurements exhibited the same deviation pattern but with different RMS values. Bad weather and moon rising during 2 measurements can have caused a significant decrease of accuracy of the method for these positions. The area with big deviations directly above the left quadrupod leg is probably the result of an adjustment error. But the astigmatism was not visible anymore. Figure **3** shows the result of the best measurements obtained after a full adjustment of all panels. Measurement was taken on Aug.15, 2011 and the surface final accuracy was 47 μm r.m.s.

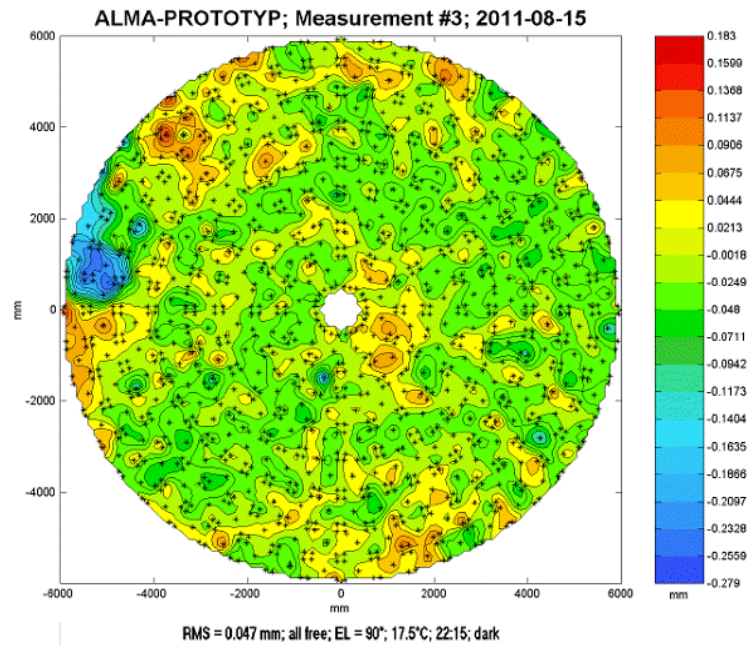

Figure 3: Photogrammetry measurement after all panels were adjusted

# 3. ANTENNA RETROFIT

Recommendations about retrofit components mainly came from VERTEX theoretical analyses and experience, but after the antenna was disassembled and components inspected, either at the VLA site or at the premises of manufacturers, we had to accept the fact that many more components would have to be refurbished or changed for the GLT to work to specifications in the Arctic environment.

## 3.1 Antenna Components to be re-used without modifications

The yoke, the receiver cabin, the invar cone and the elevation counterweights are the main antenna structural components that will be used without modifications. In fact, the insulation of all these components was stripped off for disassembly and often damaged in the dismantling process, which was not a problem since insulation would need to be redesigned and reinstalled to comply with extreme cold weather.

Panels.

The primary mirror panels are aluminum machined and will be used without modification, but each panel will be equipped with a de-ice system described in Section 3.4.

Encoders.

The antenna has 3 high-resolution absolute encoders, one on the azimuth axis, two on the elevation axis, although only one encoder is necessary for the servo loop. They were designed for operation within the temperature range of -20°C to +50°C. The encoders for elevation are located in the cabin, the temperature of which will be within +6 to +20°C, so they are compatible. The azimuth encoder is located inside the support cone, which will always be looking at the "cold" snow and will remain in the -30 to -35°C temperature range. The encoders were sent to the original manufacturer for low temperature testing and were found operational down to -35°C.

## 3.2 Antenna Components to be re-used after modifications

Carbon fiber backup structure.

The primary mirror panels are supported onto a carbon fiber reinforced polymer (CFRP) backup structure (BUS) composed of 24 box-type segments bolted together. A first detailed inspection of the segments was only possible after the entire BUS was disassembled at the VLA site in November and December 2012. Inspection was only visual and an assessment on their condition and necessary repair was needed before relocating the antenna on top of the ice sheet in Greenland, at an elevation of 3200 m and in temperatures ranging from 0°C to -73°C. The most significant damages were voids in the glue at the panel inserts, cracks in the laminate at numerous places, local delamination of the CFRP composite panels constituting the segments, some of the latter probably caused by accidental hits on the structure, others due to poor workmanship. Figure 4 illustrates some of the most worrisome damages. Repair in situ was not possible thus it was decided to send the 24 segments to AIRBORNE in the Netherlands, the manufacturer of the segments for the ALMA-NA Production antennas.

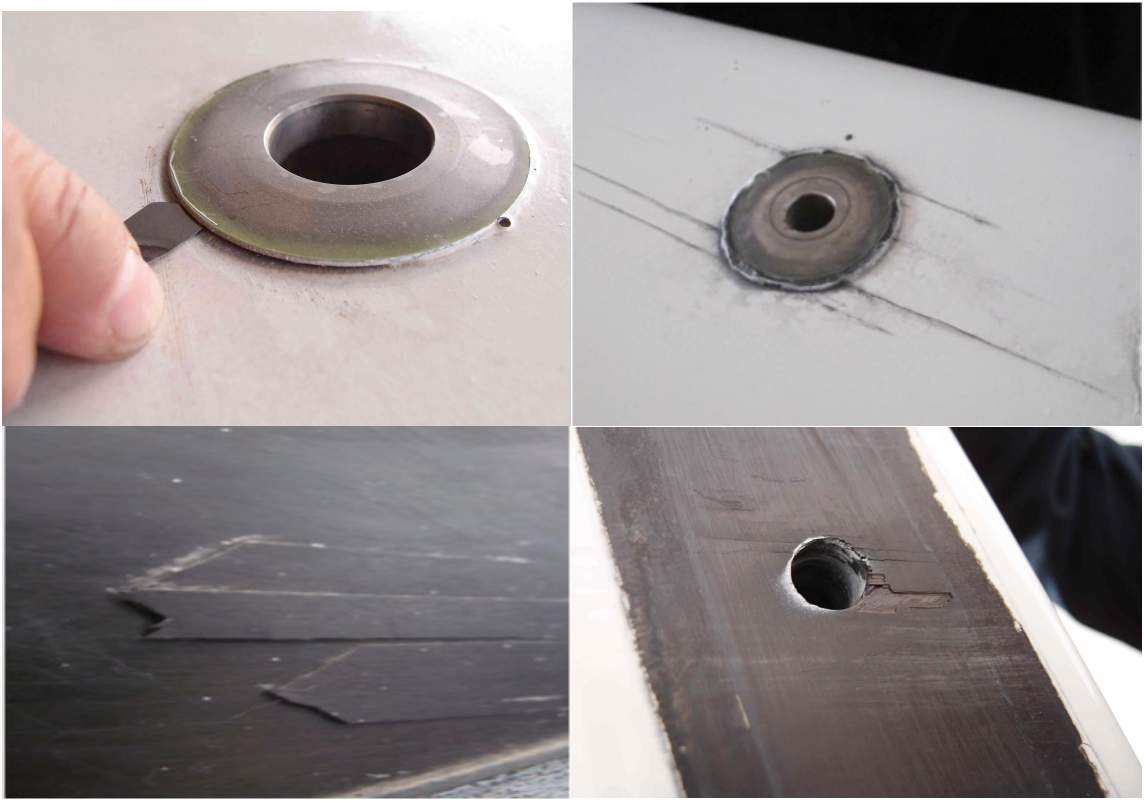

Figure 4: Damages encountered on the ALMA-NA Prototype antenna segments after disassembly at the VLA site. Voids at the panel support inserts (upper left), cracks at a quadrupod leg insert (upper right), delamination (lower left) and void and delamination (lower right)

In total we identified about 300 spots in need of attention. The large number of flaws was potential moisture ingress areas and made it inadequate for a use in an arctic environment, without being repaired.

Airborne made a very thorough inspection of the segments at their factory, and discovered more unacceptable defects. All the repairs were carried out as well as necessary modifications to be able the use of improvements made by VERTEX on the ALMA Production antennas, such as a new attachment design for the cladding panels, new grounding fixtures, but also modifications for arctic use, such as holes for gap snow cover holders & holes for the de-ice cables. All segments were then shipped from the Netherlands to Germany where the GLT team underwent training on backup structure re-assembly after the stiffness of the segments was measured. It turned out that the stiffness of the segments was higher than ALMA's and the stiffness of the assembled backup structure 10% higher than ALMA reflector mean stiffness value. The backup structure has been dismounted again and will be ready to ship before the summer of 2014.

Elevation shafts and bearings

Elevation shafts had to be disassembled, as a minor modification requiring machining was necessary. The shafts and bores in the yoke arms presented an important amount of corrosion due to the presence of water in the bores. Elevation bearings were disassembled from the antenna by PSL, the bearing manufacturer, and inspected on site after being taken apart. Given the very significant corrosion of the raceways, PSL strongly recommended refurbishing the bearings, by producing new rings and rollers. Bearings and shafts were sent to Slovakia for rework and are ready for re-installation on the antenna.

### 3.3 Antenna Components that need to be discarded and redesigned

Support cone.

The telescope cannot be supported directly on the ice sheet, but needs an interface structure on top of the ice foundation. Figure **5** is a model of the GLT on its support. The GLT team is designing the Spaceframe, similar in concept to the structure designed to support the South Pole Telescope (SPT) [5]. Figure 6 is a computer model of the support structure. The goal of the spaceframe is to distribute the load of the telescope on the foundation and delay snow accumulation and drifts around the telescope. To better distribute the loads of the antenna, the support cone itself was redesigned with an adjustable 6-point interface to the spaceframe. The features of this new support cone are: possibility of leveling the antenna at this 6-point interface when needed, open bottom so that the cone and inside components are always at the temperature of the snow (-30 to -35°C), material is low temperature steel, support structures for encoder, and increased stiffness due to the full cone geometry connecting to 6 points instead of a truncated cone structure as shown on Figure 7.

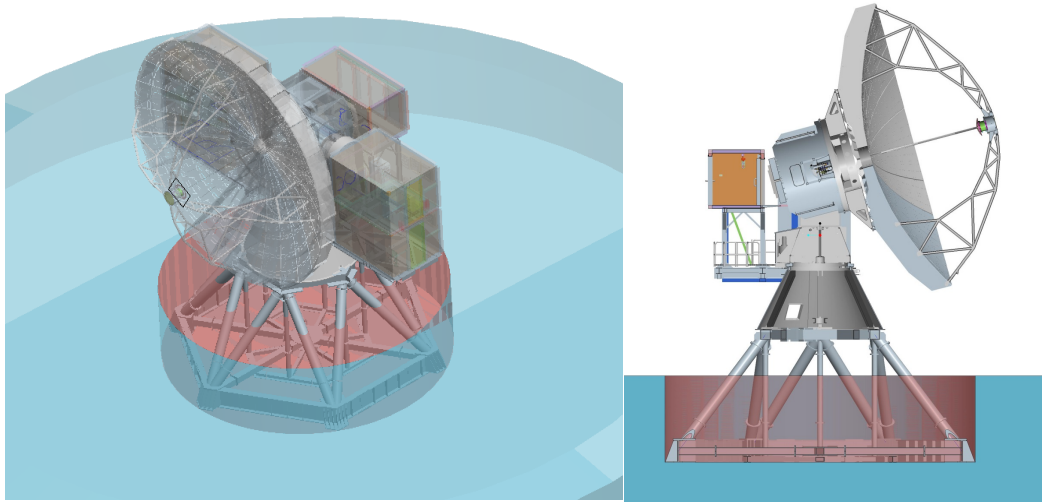

Figure 5: Model of the GLT on the space frame partially embedded in the snow foundation (design of space frame and snow pad in progress). Isometric view showing all 4 side containers and receiver transfer container on the back (Left); sectional side view (Right)

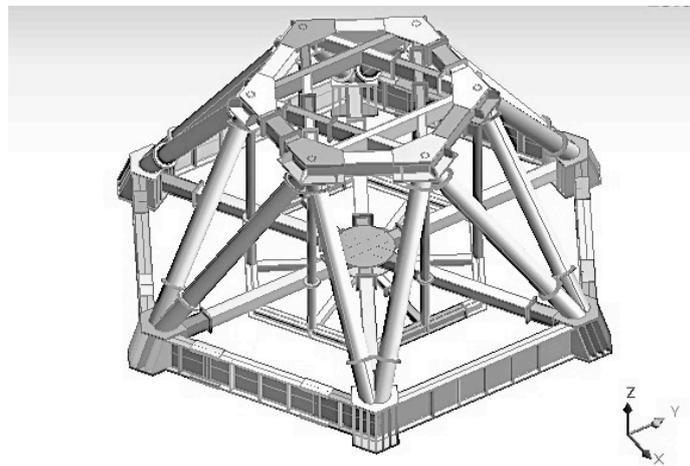

Figure 6: Model of the spaceframe to support the GLT on the ice sheet

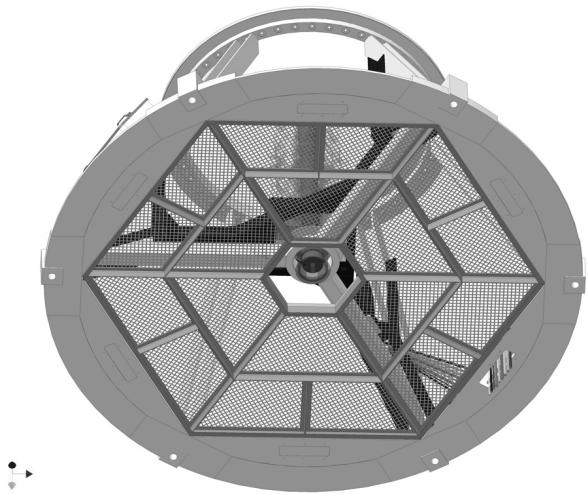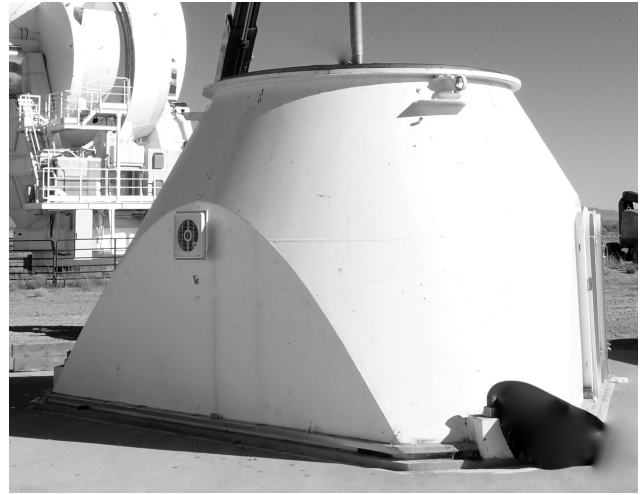

Figure 7: New support cone for the GLT showing bottom open to snow environment, full conical shape with 6 points support (left), ALMA-NA prototype antenna support cone, truncated, 3 support point (right)

Azimuth bearing

It was not possible to inspect the raceways on site, so the bearing was shipped to ROTHE ERDE, the original manufacturer in Germany, for inspection. ROTHE ERDE strongly recommended making a new nose ring because of significant wear in the gearing and pitting in both flanges of teeth, probably the result of lack of lubrication. It was decided to purchase a new bearing fully compliant to the Greenland site operation, given the minimal cost difference between a new bearing and refurbishing the old one. The new bearing is guaranteed by the manufacturer to work in the cold environment. The bearing passed the factory acceptance tests in April at the plant, it was fully lubricated with special low temperature grease, friction torques under symmetrical and asymmetrical load were measured, as well as bearing wobble.

HVAC system & insulation

The HVAC system was not functional during the re-test and in any case a system designed to work in the Atacama Desert could not be used in Greenland. The redesign of the HVAC system and insulation does not only involve maintaining the receiver cabin at a particular range of temperature during operation and survival conditions, but it directly impacts the pointing of the telescope and its surface accuracy. This is the reason why we contracted VERTEX to provide thermal analyses of the retrofitted telescope and specifications on HVAC components and insulation.

Servo system

The servo system of the Prototype antenna needed to be replaced to comply with GLT specifications. Most of the components were either obsolete (ACU or PLC), would not work at low temperature (elevation and azimuth motors, limit switches and all cabling) or with low reliability after more than 10 years of existence, as was discovered during the re-commissioning performed by VERTEX in 2011. New motors were procured and tested at low temperature. The new servo is based on ALMA Production antenna system. The factory acceptance test of the servo system was successfully completed in April at VERTEX. The ALMA-NA prototype antenna only used one tiltmeter on the azimuth bearing, operational in the temperature range -20°C to +40°C. The GLT will use 2 high accuracy tiltmeters, one over the azimuth bearing, and the other on the support cone. New tiltmeters have been procured which will work down to -35°C. The new tiltmeters have been tested at the vendor's premises and work fine at temperatures around -50°C. They are currently being calibrated. There was no alternative than ordering new gearboxes both for elevation (4) and azimuth (2) as the design and material used for the ALMA-NA prototype antenna were not compliant with the GLT operational and survival temperatures.

Quadrupod

The original quadrupod of the ALMA-NA prototype antenna had been damaged twice during the ALMA testing period at the VLA, and repaired. It was however impossible to assess whether the components used for the repair work could withstand the harsh environment of the Greenland site. Furthermore, after the reflector was removed from the antenna and the quadrupod taken apart, inspection showed another outstanding damage. It was decided to order a new quadrupod from VERTEX, identical to the ALMA Production antenna.

Hexapod

The PI hexapod installed on the ALMA Prototype antenna could not be used for Greenland operations and there was no possibility to refurbish it either. A new hexapod is under procurement at ADS, fully compliant with the GLT specifications. This hexapod will also have a nutating function and will be designed to allow holography. Design is in progress and should be completed by the end of 2014.

### 3.4 New components

Primary mirror panel de-ice system.

Ice formation has to be avoided on the reflector primary mirror panels at Summit Station. Given the successful experience of the South Pole Telescope [5] in that regard, we are designing a de-ice system to raise the temperature of the panels from 2 to 5K above ambient, by attaching heating pads below each panel. SPT runs the system continuously raising the temperature 1-2 K above ambient. We are designing a system with a maximum power density of 300 W/m$^2$, 12 antenna radial zones with quadrant heating mode control, 48 panels equipped with temperature sensors (18% of the total number of panels) and air temperature sensors installed on the antenna. The design is completed.

Antenna containers

Electronic equipment of the ALMA-NA Prototype antenna, apart from the receiver cabin, was housed in the support cone, and on a platform supported by two I-beams attached under the yoke structure. This arrangement had to be fully redesigned for the GLT, and the plan is to accommodate all the electronics, power distribution, servo cabinet, UPS, 120 and 480V load centers, receiver control computer, compressors, chillers, hexapod control system, and de-ice system in the four containers. The antenna containers are in their final design phase and a snapshot can be seen below in Figure 8. The main challenge of the design of these new enclosures is to minimize the resulting overturning moment on the foundation. The design may look similar to the APEX antenna [6], but the GLT remains a Cassegrain antenna and the overall weight on the two I-beams connecting the container system to the antenna is only 20 000kg, amounting for about 60% of the additional structure of APEX.

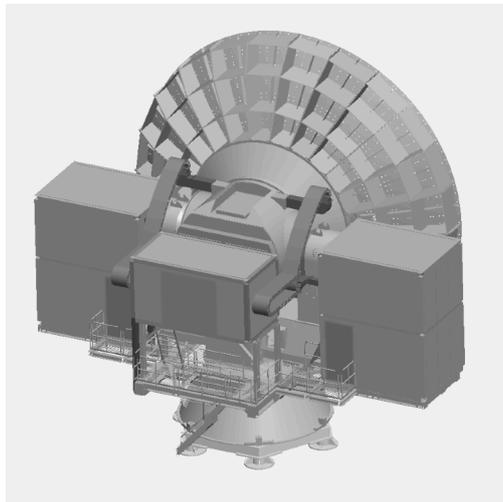

Figure 8: GLT computer model showing the 2 lateral containers at each side of the antenna, the receiver transfer container at the back and access platforms. (Design in progress)

## 4. FUTURE PLANS

The GLT components are spread all around the world at this time: New Mexico, Germany and Taiwan and the plan is to ship all parts ready early summer of 2014 to a site of the US East Coast. The US Army Cold Regions Research and Engineering Laboratory (CRREL) in Hanover, NH, has agreed to host us on their campus for the GLT Team to test-fit the antenna. CRREL has a considerable experience in designing structures for Arctic and Antarctic regions, and will be involved in the design of the foundation for the GLT at the Greenland site.

Only one boat a year leaves Norfolk Naval Base for Thule in June, and the telescope components will be transported from Hanover, New Hampshire to Norfolk.

An important milestone for the GLT is to take part in the detection of the Black hole shadow in M87 with ALMA phased-up project, as highlighted by Inoue et al [1]. Therefore the GLT needs to be re-assembled and operational in Greenland by 2016-2017.

We are monitoring the opacity at Summit Station in Greenland with a 225 GHz radiometer, and Martin-Cocher et al [7] will present the results of measurements at this conference. Recent work showed that VLBI observations towards M87 could also be carried out at Thule, at 230 and possibly 350GHz, as the Greenland site is expected to be ready in 2018 for final commissioning of the GLT.